\documentclass[12pt,a4paper]{article}
\usepackage{graphicx}
\usepackage[cp1251]{inputenc}
\usepackage{amssymb, amsfonts, amsmath}
\date{}
\usepackage{amscd}
\begin{document}

\title{Exact solutions of Maxwell vacuum equations in null homogeneous Petrov spaces with solvable motions groups}

 \author{V.V. Obukhov}

 \maketitle


            \quad

Tomsk State Pedagogical University (TSPU),  Institute, of Scietific Research and

Development, 60 Kievskaya St., Tomsk, 634041, Russia;

 e.mail: obukhov@tspu.ru.

 \quad

 Tomsk State University of Control Systems and Radio Electronics (TUSUR),

 Laboratory for Theoretical Cosmology, International Center of Gravity and Cosmos,

 36, Lenin Avenue, Tomsk, 634050, Russia

\maketitle

\section{Introduction}

%
%
%

In this article  Riemannian spaces on which three-parameter groups of motions  $\tilde{G}(N)$ act simply transitively are considerate. The following  notation and symbols are used in the text. The symbol $W_4$ denotes a Riemannian space with Lorentz signature. Non null  hypersurfaces of $W_4$ are denoted by the symbol $\tilde{V}_3$. The symbol $\tilde{V}^*_3$ denotes null hypersurfaces. The group $\tilde{G}(N)$ is related to the group $G(N)$ in the Bianchi classification by the homomorphism
$$
G(N) \mapsto \tilde{G}(N).
$$
Here and below, $N$ denotes the number of the group $G_3(N)$ in the Bianchi classification (from $I$ to $IX$, see \cite{1n} - \cite{4n}). The invariant space of the group $G_3(N)$ is a three-dimensional Riemannian manifold $V_3$ on which the group $G_3(N)$ acts simply transitively.
 The space $W_4$ is often called a homogeneous spacetime of type $N$ according to the Bianchi classification if the invariant space of the group $\tilde{G}_3(N)$ is a space-like hypersurface $\tilde{V}_3$. This definition of a homogeneous spacetime can be generalized.

\begin{enumerate}

\item

A Riemannian space $W_4$ with Lorentz signature is called a non-null homogeneous space of type $W_4(N)$ according to Bianchi (or a non-zero homogeneous spacetime of type $N$ according to Bianchi's classification) if the invariant space of the group $\tilde{G}_3(N)$ is the non-null hypersurface $\tilde{V}_3$.

\item

A Riemannian space $W^*_4$ with Lorentz signature is called a null homogeneous space of type $W^*_4(N)$ according to Bianchi (or a null homogeneous spacetime of type $N$ according to Bianchi's classification), if  the group of motions $\tilde{G}_3(N)$ acts on the null hypersurface $\tilde{V}^*_3$.

\end{enumerate}

In the space $W_4(N)$, we can introduce a non null semi-geodesic coordinate system [$u^i$] in which the hypersurface $\tilde{V}_3$ is defined by equation $u^0 = const$ (see, for example, \cite{3n}-\cite{5n}). Obviously, the group of motions $G_3(N)$, which acts simply transitively in the invariant space $V_3(N)$, and the group of motions $\tilde{G}_3(N)$ are related by an isomorphism:
$$
G_3(N) \leftrightarrow \tilde{G}_3(N),
$$
since the generators of these groups are defined by the same Killing equations, and the components of the Killing vector fields in the space $W_4(N)$ do not depend on the variable $u^0$
(from now on, we will use the same symbols to denote isomorphic groups).

The geometry of the space $W_4(N)$ is determined by the geometry of the space $V_3(N)$. Due to the isomorphism of the groups $G_3(N)$ and $\tilde{G}_3(N)$, the components of the metric tensor of the hypersurface $\tilde{V}_3(N)$ coincide with the components of the metric tensor of the space $V_3(N)$, in which the integration constants of the Killing equations are considered as functions of the parameter $u^0$. In what follows, we will denote these non-null homogeneous spaces by the symbols $V_4(N)$.

The groups of motions $\tilde{G}^*_3(N)$  (acting simply transitively on the null hypersurfaces $\tilde{V}^*_3$ of the space $W^*_4$) are in general case related by a homomorphism to the groups $G_3(N)$ (
acting on the invariant spaces $V_3$):
$$
G_3(N)\rightarrow \tilde{G}^*_3(N).
$$
Let us consider this statement in more detail.
One can introduce in the space $W^*_4$ a null semi-geodetic coordinate system [$u^i$], in which the null transitivity hypersurfaces $\tilde{V}^*_3(N)$ of the motions group $\tilde{G}^*_3(N)$ are given by the equation $u^0=const$. Admissible transformations of the coordinate system [$u^i$] must ensure the preservation of this equation, as well as the independence of the components of the Killing vector fields from the variable $u^1$. These restrictions are more stringent than in the case of a non-null transitivity hypersurface (see \cite{3n}-\cite{5n}).
Therefore, the classification of solutions of the Killing equations with respect to the group of admissible transformations leads to the fact that each group $G_3$ can generate up to three nonequivalent (with respect to the group of admissible transformations of semi-geodetic coordinates) motions groups  $G_3(N)$, acting in nonequivalent invariant spaces $V_3(N)$. In the case where the number $n$ of these nonequivalent groups of motions is greater than $1$, we will provide each nonequivalent group and the corresponding invariant space with an additional index $n$:
$$
G_3(N)_{n}, \quad V_3(N)_{n}.
$$
Some of the solvable groups $G_3(N)_{n}$ (the groups $\tilde{G}_3(II)_n, \tilde{G}_3(III)_n, \tilde{G}_3(V)_n$) generate one-parameter families of groups of motions $\tilde{G}_3(N)_{n}$ of the spaces $W^*_4(N)$ . The above-mentioned homomorphism holds for the groups belonging to each such family. In what follows, we will consider only those null homogeneous spaces of type $W^*_4(N)$ for which  group operators  do not depend on the variable $u^0$. We will denote such spaces by the symbols $V^*_4(N)_n$. The motions groups of the spaces $V^*_4(N)_n$ are isomorphic to the groups of motions $G_3(N)_{n}$ of the invariant spaces $V_3(N)_n$:
$$
G_3(N)_{n}\leftrightarrow \tilde{G}_3(N)_{n}.
$$
As in the case of non-null homogeneous spaces $V_4(N)$, the geometry of the null homogeneous spaces $V^*_4(N)_n$ is determined by the geometry of the invariant spaces $V_3(N)_n$.

In the books \cite{3n}-\cite{4n} (see \cite{5n}-\cite{6n} too), a complete classification of all homogeneous spaces of type $N$ according to Bianchi was provided. The components of all nonequivalent metric tensors $g_{ij}$ and the components of all nonequivalent Killing vector fields $\xi_a^i$ were found explicitly. Therefore, it is proposed to call the spaces $V_4(N)$ and $V^*_4(N)_{n}$ homogeneous Petrov spaces.

Besides homogeneous spaces of type $N$, another class of Riemannian spaces is known that admit complete sets of Killing fields, and hence symmetry operators for the equations of motion of a test particle. The complete sets consist of triplets of independent mutually commuting Killing fields (vector and/or tensor fields of no higher rank than second). The symmetry operators generated by these fields form complete sets that play an important role in classical and quantum mechanics. The corresponding spaces are called Stackel spaces (see \cite{7n}-\cite{18n}). The classification problem for Stackel spaces was finally solved by V.N. Shapovalov (see \cite{11n}-\cite{13n}). The components of all nonequivalent metric tensors $g_{ij}$ and the components of all nonequivalent Killing vector fields $\xi_a^i$ were found explicitly.

 Stackel spaces and homogeneous Petrov spaces also provide methods for the exact integration of the  motion equations of test particles. For a Stackel space, this is the method of complete separation of variables (the method of commutative integration). For homogeneous Petrov spaces, this is the method of non-commutative integration (see \cite{19n}, \cite{20n}. Therefore, these classes of spaces have wide application in gravity theory, primarily for solving the exact integration problems of the  motion equations and field equations. These equations in privileged  coordinates  are reduced to systems of ordinary differential equations ( for the cases of simply transitive four-parameter motions groups $G_4$
the equations are reduced  to systems of algebraic equations; see \cite{6n}).

It should be noted that many of the best-known exact solutions of Einstein equations belong to the classes of homogeneous Petrov or Stackel spaces.
The development of exact integration methods has been carried out in many papers. For example, in \cite{20n} and \cite{21n} the non-commutative integration method was applied to obtain exact solutions of Hamilton-Jacobi equation and Klein-Gordon-Fock equations. In \cite{22n} - \cite{24n} the method was generalized for the Dirac equations and for Yang-Mills gauge fields.

The set of non null homogeneous Petrov spaces includes, for example, spaces with multi-transitive groups of motions. These spaces have long attracted the attention of specialists studying the problem of exact integration of Einstein equations (see the book \cite{6nn} and the bibliography therein, as well as the articles \cite{24nn} - \cite{26nn}).

Such spaces include, for example, spatially homogeneous cosmological models with an ideal fluid (see \cite{25n} - \cite{27n}). These models also belong to the class of Stackel spaces, which makes them more effective in studying the motion equations of a matter, which is used in various theories of gravity (see, for example, \cite{29n} - \cite{32n}). The set of non-null homogeneous spaces of the type $V_4(N)$, possessing Stackel metrics, was also studied in \cite{33n} - \cite{38n}. Non-null homogeneous Petrov spaces have also been fairly well studied. Various problems of gravity theory have been considered in numerous papers, including the study of exact solutions of Einstein equations (see, for example, \cite{38nn}).

The problem of exact solutions of Einstein equations for the case where $G_3$ are maximal groups of motions was considered in the papers \cite{39n}-\cite{42n} for spaces $V_4(I)$ (a detailed bibliography can be found in the book \cite{6nn}).
A complete classification of electro vacuum spaces with cosmological term for Petrov space of type $V_4(I)$ was carried out in the works \cite{43n}-\cite{44n}.

At the same time, the class of null homogeneous Petrov spaces of type $V^*_4(N)$ has been little studied. In the papers \cite{45n}, \cite{45nn}, \cite{46n} these spaces were considered for the case where the complete set of the space includes a null Killing vector field. It was established that vacuum fields, Einstein-Maxwell fields, and pure radiation fields that admit a simply transitive group $G_3(I)$ with null operator are plane  waves.

Using a canonical frame in non-null homogeneous Petrov spaces allows one simplify the field equations. This is of great importance when solving problems in mathematical physics. Due to the analogy between non-null and null homogeneous Petrov spaces, the same possibility exists for null homogeneous Petrov spaces. In present paper, this is demonstrated using the example of the Maxwell vacuum equations. The problem of classifying of exact solutions of vacuum Maxwell  equations in $V^*_4(N)$ spaces for the case of solvable motions groups is solved. This problem should be considerate as a stage in solving the general classification problem for spaces with complete sets (for Stackel spaces and for homogeneous Petrov spaces). Let us outline the main stages of the general classification.

\begin{enumerate}

\item
Finding all  classes  of equivalence for the spaces with complete sets. Simultaneously, finding all complete sets of integrals of motion of free Hamilton-Jacobi equation for each space.

\item
Finding all nonequivalent electromagnetic vector potentials in which the motion equations  admit integrals of motion defined by complete sets.
The classification should be constructed separately for the Hamilton-Jacobi, Klein-Gordon-Fock and Dirac-Fock equations (including those with a Yang-Mills field).

\item

Finding all nonequivalent exact solutions of vacuum  Einstein-Maxwell equations for the spaces and invariant electromagnetic fields with complete sets.

\end{enumerate}

By now, the classification problems related to the first stage have been completely solved.
For Stackel spaces, they were solved in papers \cite{5n}-\cite{17n}. For homogeneous Petrov spaces, the classification was carried out in Petrov's books \cite{3n}, \cite{4n}, (see also \cite{5n}).

The second-stage problems were solved for homogeneous Petrov spaces (see articles \cite{21n}, \cite{6n} and the bibliography in it). For  the  flat space-time (a special case of the Stackel space) the problems were solved for  Hamilton-Jacobi and Klein-Gordon-Fock equations (see \cite{51n}-\cite{53n}). For other Stackel space the classification problem was solved for the Hamilton-Jacobi equation in papers \cite{14n}, \cite{15n} (see also bibliography therein).

The problems of the third stage were solved for the electro vacuum Stackel spaces (see \cite{17n}, \cite{41n}-\cite{43n}). The first part of the classification problem of non-null electro vacuum homogeneous Petrov spaces was solved in papers \cite{56n}-\cite{58n}. All non-equivalent exact solutions of the Maxwell vacuum equations for the non-null homogeneous Petrov spaces were found.

In this paper, the problem of classifying exact solutions of the Maxwell vacuum equations for the null homogeneous Petrov spaces with solvable motions groups ($G_3(I)-G_3(VII)$ is completely solved.

The results of the classification of null homogeneous Petrov spaces with non-solvable motions groups ($G_3(VIII)-G_3(IX)$) will be presented in another paper.

The paper is structured as follows.

\begin{enumerate}

\item The second section provides all necessary definitions and notations used in the text.

\item In the third section, Maxwell equations are obtained for null homogeneous Petrov spaces. The calculations are performed in a canonical frame. The resulting equations contain only the non-holonomic components of the metric tensors and of the vector potentials of invariant electromagnetic fields.

\item The fourth section presents all nonequivalent solutions of Maxwell vacuum equations in null homogeneous Petrov spaces with solvable groups of motions. For each group, all nonequivalent exact solutions of Maxwell vacuum equations and all nonequivalent sets of vector fields of canonical frame are presented.

\end{enumerate}

\section{HOMOGENEOUS PETROV SPACES}

First of all, one has to clarify the notations. Throughout the text, the following notations  used:
$$ i,j,k,r \div 1,2,3,0;\quad p,q \div 2,3; \quad \alpha ,\beta ,\gamma ,\delta ... \div 1,2,3;\quad a,b,c,d\div 1,2,3.$$
The fourth coordinate of the  semi-geodetic coordinate system  $\left\{u^i\right\}$ in spaces $V_4(N)$, $V^*_4(N)$ is supplied with the index $i=0$ ($u^0$), and the numbering of coordinate indices always starts with one and ends with zero (when indices $i,j,k,l$ are used). Greek letters denote the coordinate indices of the coordinate system $\left\{u^{\alpha} \right\}$, related to the invariant space $G_{3} \left(N\right)$. The Latin letters $a, b, c$ denote indices of structure constants, non-holonomic coordinates of vector and tensor fields associated with the canonical frame $\zeta^\alpha_a$.  In the case where the indices $p,q$ are placed in brackets ((p), (q)), they number the vector fields in the canonical frame or in the complete  set.

The geometry of the homogeneous Petrov space is determined by the geometry of the invariant space $V_3(N)$ of the group $G_{3} \left(N\right)$. Finite motions in  space $V_3(N)$ are defined by the generators:
\begin{equation}\label{n1}
Y_a = \zeta^\alpha_a p_a, \quad [Y_a, Y_b]= C^c_{ab}Y_c.
\end{equation}
The vector fields $\zeta^\alpha_a$ form a canonical frame in the space  $V_3(N)$  in which
the components of the metric tensor $G^{\alpha\beta}$ of the space $V_3(N)$ can be represented in the form (see \cite{59n}):
\begin{equation}\label{n2}
G^{\alpha\beta} = \zeta^\alpha_a\zeta^\beta_b \eta^{ab}, \quad \eta^{ab}=const
\end{equation}
Infinitesimal motions of the space $V_3(N)$ are defined by the generators:
\begin{equation}\label{n3}
X_a = \xi^\alpha_a p_a, \quad [X_a, X_b]= \tilde{C}^c_{ab}X_c,
\end{equation}
where $\xi^\alpha_a$ are the components of the Killing vector fields of space $V_3(N)$, satisfying the Killing equations:
\begin{equation} \label{n4}
G^{\alpha\beta}_{,\gamma}\xi_{a}^{\gamma} = G^{\alpha\gamma}\xi^{\beta}_{a,\gamma} + G^{\beta\gamma}\xi^{\alpha}_{a,\gamma}.
\end{equation}
The Killing vector fields $\xi^{\alpha}_a$ are connected with the vector fields of the canonical frame $\zeta_a^{\alpha}$ by sets of equations (see \cite{59n}):
\begin{equation} \label{n5}
\zeta_{a,\beta }^{\alpha } =\xi_{\beta }^b \xi_{b,\gamma }^{\alpha}\zeta_a^{\gamma} \Leftrightarrow
\xi_{a,\beta}^{\alpha} =\zeta_\beta^b \zeta_{b,\gamma}^{\alpha } \xi_a^{\gamma }.
\end{equation}
Here
\begin{equation} \label{n6}
\xi_{\beta }^b \xi_{b}^{\alpha} = \zeta_{\beta }^b \zeta_{b}^{\alpha} =\delta_\beta^{\alpha}.
\end{equation}
In a non null semigeodesic coordinate system $[u^i]$, the metric tensor of the space $V_4(N)$ has the form:
\begin{equation} \label{n7}
g^{ij}(u^i) =\left(\begin{array}{cccc} {} & {} & {} & {0} \\ {} & {g^{\alpha \beta }} & {} & {0} \\ {} & {} & {} & {0} \\ {0} & {0} & {0} & {\varepsilon} \end{array}\right), \quad \varepsilon = \pm 1, \quad \left\|g^{ij}\right \| < 0.
\end{equation}
The components of $g^{\alpha\beta}$ obey the same Killing equations \eqref{n4} as the metric tensor $G^{\alpha\beta}$ of the invariant space $V_3(N)$. Therefore, the tensor $g^{ij}$ of the space $V_4(n)$ can be represented as:
\begin{equation} \label{n8}
g^{ij} = \ell^i_{(l)} \ell^j_{(k)}\eta^{(lk)}(u^0),
\end{equation}
where the tetrad $\ell^i_{(j)}$ and the matrix $\eta^{(ij)}$ have the form:
$$
\ell^i_{(j)}= \delta^i_0\delta_{(j)}^{(0)} + \delta^i_{\alpha}\delta_{(j)}^{(a)}\zeta^{\alpha}_{(a)}, \quad \eta^{(ij)} =\left(\begin{array}{cccc} {} & {} & {} & {0} \\ {} & {\eta^{ab}(u^0)} & {} & {0} \\ {} & {} & {} & {0} \\ {0} & {0} & {0} & {\varepsilon} \end{array}\right) \quad (\varepsilon\left\|\eta^{ab}\right \| < 0).
$$
 The use of the canonical tetrad $\ell^i_{(j)}$ makes the spaces $V_4(N)$ particularly convenient for solving the Maxwell, Einstein, and Einstein-Maxwell vacuum equations, since the tetrad components of the Ricci tensor of the space $V_4(N)$ contain only the structure constants $C^a_{bc}$ and the functions $\eta^{ab}, \dot{\eta}^{ab},\ddot{\eta}^{ab}$ (see \cite{59n}).
 The dots denote derivatives with respect to the variable $u^0$. In an invariant electromagnetic field, the tetrad components of the tensor $F^{ij}_{;j}$, in addition to these quantities, also contain the tetrad components of the electromagnetic potential $a_{(j)}=\ell^i_{(j)}A_i$ and the functions $\dot{a_b}, \ddot{a}_b$ (see\cite{56n}). Throughout the text, the gradient transformation of the potential $A_i$ ensures the condition:
$$A_0=0.$$
The same circumstance one has  for the null homogeneous Petrov space $V^*_4(N)$. In Petrov's books (\cite{3n}, \cite{4n}) it was shown that the metric tensor of a spacetime $W^*_4(N)$ admitting a simply
transitive group of motions, acting on the null hypersurface $\tilde{V}^*_3(N)$, can be represented (in the null semi-geodetic coordinate system $[u^i]$) as follows:
\begin{equation} \label{n9}
g_{ij}(u^i) =\left(\begin{array}{cccc} {0} & {0} & {0} & {1} \\ {0} & {} & {} \\ {0} & {} & {g_{kl}} \\ {1} & {} & {} & {}\end{array}\right)\quad (k, l = 0, 2, 3), \quad g^{ij}(u^i) =\left(\begin{array}{cccc} {} & {} & {} & {1} \\ {} & {g^{\alpha \beta}} & {} & {0} \\ {} & {} & {} & {0} \\ {1} & {0} & {0} & {0} \end{array}\right).
\end{equation}
The form \eqref{n9} is equivalent to that used in Petrov's books. The Killing vector field satisfies the conditions:
\begin{equation} \label{n10}
\xi^\alpha_{(a),1} = 0, \quad \xi^i_{(j)} = \delta^i_{\alpha}\delta_{(j)}^{(a)}\xi^{\alpha}_{(a)}.
\end{equation}
By conditions \eqref{n9} \eqref{n10}, the Killing equations  include the functions $g^{\alpha \beta }, \xi^{\alpha}_{(a)}$. If we additionally require that the components of the Killing vector fields be independent of the variable $u^0$:
\begin{equation} \label{n11}
\xi^{\alpha}_{(a),0}=0,
\end{equation}
then $g^{\alpha \beta }$ will be the components of the metric tensor of the invariant space $ V_3(N)$, which can be represented in the same manner as \eqref{n2}:\quad $$ g^{\alpha\beta} = \zeta^\alpha_a\zeta^\beta_b \eta^{ab}(u^0).$$
The vector field $\zeta^\alpha_a$ can be found from the system of equations \eqref{n5}.
In this case, the space $W^*_4(N)$ has been called the null homogeneous Petrov space of type $ V^*_4(N)$.
The constraint \eqref{n10} naturally selects from the set of all Petrov spaces a set of null homogeneous spaces whose geometry is completely determined by the geometry of the invariant space $V_3(N)$. In the null homogeneous Petrov spaces $V^*_4(N)$ the non-holonomic components of the Ricci tensor and the tensor $F^{ik}_{,k}$ in the canonical frame $\zeta^\alpha_{(a)}$ contain only the structure constants $C^a_{bc}$ and the functions \quad $\eta^{ab}, \dot{\eta}^{ab},\ddot{\eta}^{ab}, a_{(a)}, \dot{a_b}, \ddot{a}_b$ in the same manner as in the non null homogeneous Petrov spaces $V_4(N)$. In the present paper, this statement is proved for vacuum Maxwell equations.

\section{Maxwell vacuum equations in the null homogeneous Petrov spaces}

In the article \cite{6n} (see bibliography therein) it was proven that in homogeneous Petrov spaces the Hamilton-Jacobi and Klein-Gordon-Fock equations for a charged test particle admit integrals of motion that are linear in momenta if the electromagnetic field is invariant under the action of the motions group $G_3(N)$. The invariance condition can be represented as follows:
$$
(\xi_a^\alpha A_\alpha)_{,\beta}\xi_b^\beta=\tilde{C}^c_{ba}A_\alpha\xi_c^\alpha.
$$
The solution of this equation has form:
\begin{equation}\label{n12}
A_\alpha = \zeta_\alpha^a a_a(u^0), \quad A_0=0.
\end{equation}
Let us consider the Maxwell vacuum equations:
\begin{equation}\label{n13}
\frac{1}{\sqrt{g}}(\sqrt{g} F^{ij})_{j}=0 \quad (g=-det|g_{kl}|) \quad \Rightarrow F^{ij}_{,j} -\frac{1}{2g} g_{,j} F^{ij} =0.
\end{equation}
It can show that from equations \eqref{n5} it follows:
\begin{equation}\label{n14}
\zeta_{(1)}^\alpha =\delta^1_a, \quad \zeta_1^{(a)} =\delta_1^a.
\end{equation}
From \eqref{n5} it follows:
$$
\zeta^1_a = C^1_a + C^2_a\varphi^1 + C^3_a\omega^1, \quad \zeta^2_a = C^2_a\varphi^2 + C^3_a\omega^2, \quad \zeta^3_a = C^2_a\varphi^3 + C^3_a\omega^3,
$$
where $\varphi^\alpha, \omega^\alpha$ are some functions depending on variables $u^1, u^2, u^3;$ $ C^\alpha_a$ are integration const.

\noindent
Then it follows:
$$Y_a = C^1_a p_1 + C^2_a(\varphi^\alpha p_\alpha) +  C^3_a (\omega^\alpha p_\alpha).$$
Then operators $Y_a$ can be chosen in the form:
$$
Y_1 =p_1, \quad Y_2 =\varphi^\alpha p_\alpha, \quad Y_3 =\omega^\alpha p_\alpha.
$$
Components $g^{pq}$ can be represented in the form:
\begin{equation}\label{n15}
g^{pq} =G^{pq} = \zeta^{p}_{(p')} \zeta^{q}_{(q')}\eta^{p'q'}(u^0), \quad $$
$$G_{pq} = \zeta^{(p')}_{p} \zeta^{(q')}_{q}\eta_{p'q'}, \quad G^{pq'}G_{qq'}= \eta^{pq'}\eta_{qq'} = \zeta^{(p')}_{q} \zeta_{(p')}^{p}=\delta^p_q.
\end{equation}
Let us represent the commutation relations \eqref{n1} in form:
\begin{equation}\label{n16}
\zeta^\alpha_{a|b}-\zeta^\alpha_{b|a}=C^c_{ba}\zeta^\alpha_c,
\end{equation}
where the notation is used:
$$
\zeta^\alpha_{a|b}=\zeta^\beta_{b}\zeta^\alpha_{a,\beta}.
$$
One can show:
\begin{equation}\label{n17}
\zeta^\alpha_{a,\alpha}+\frac{g_{|a}}{2g}= C_a \quad (C_a = C_{pa}^{p}).
\end{equation}
The equality:
\begin{equation}\label{n18}
\frac{g_{|a}}{g}=-g_{ij}g^{ij}_{|a}.
\end{equation}
(see \cite{59n})
Due to account \eqref{n14} one has:
$$
\det|g^{ij}| = -\det|G^{pq}| =-\exp2(\sigma_0\sigma).
$$
Here $$\exp\sigma = \det|\zeta^{p}_{q}|, \quad \exp2\sigma_0 = \det|\eta^{pq}|).$$
From the other hand:
$$
\frac{g_{|a}}{2g}=-\frac{G_{|a}}{2G}= -\zeta^{q}_{(p)|a}\zeta_{q}^{(p)} \quad (G= \det{G^{pq}}) \Rightarrow
$$
\begin{equation}\label{n19}
\zeta^\alpha_{(a),\alpha} +\frac{g_{|a}}{2g}= \zeta^{(b)}_\beta \zeta^\beta_{(a)|b} +\frac{g_{|a}}{2g} = \zeta^{(p)}_q \zeta^q_{(p)|a} + C_{pa}^{p} + \frac{G_{|a}}{2G}
 = C_a .
\end{equation}
(equation \quad $\zeta^1_{a,1}=0$ is used).  The equality \eqref{n16} is proven.
As is known (see \cite{59n}), in non-null homogeneous spaces, the non-holonomic components of Ricci tensor:
$$
R^{ab}=\zeta_i^a\zeta_j^b R^{ij}, \quad R^{0b}=\zeta_i^b R^{0i}, \quad R^{00}
$$
in the  canonical frame $\zeta_{a}^{\alpha}$ depend only on the variable $u^0$. These components contain only the functions $\eta^{ab}, \dot{\eta}^{ab}, \ddot{\eta}^{ab} $ and the structural constants $C^c_{ab}$.
Non-holonomic components of the tensor $F^a=\zeta_i^a F^{ij}_{;j}$
contain the functions $a_b, \dot{a}_b, \ddot{a}_b$.
Let us show that for $F^a$ this is also true in null homogeneous Petrov spaces.

To obtain non-holonomic components of the tensor $F^a$ in the canonical frame $\zeta^\alpha_a$ one has to use the relations \eqref{n12}, \eqref{n14} \eqref{n17}.
The non-holonomic components $F_{ab}, F_{0b}$ of the electromagnetic field tensor in the canonical frame $\zeta_a^\alpha$ of the null homogeneous Petrov space will be denoted as follows:
$$
F_{ab}=\zeta_a^\alpha\zeta_b^\beta F_{\alpha\beta} =C^c_{ba}\alpha_c, \quad F_{0a}=\zeta_a^\alpha F_{0\alpha}= \dot{\alpha}_a.
$$
The indices $a,b$ are raised and lowered using the non-holonomic components of the metric tensor \quad $\eta^{ab}, \eta_{ac}, \quad (\eta_{ac}\eta^{ac} =\delta^b_a)$ of the space $ V_3(N)$:
\begin{equation}\label{n20}
F^{ab}=\eta^{aa_1}\eta^{bb_1}F_{a_1b_1}, \quad F^{a}_{\quad b}=\eta^{ac}F_{cb}.
\end{equation}
Let us introduce the functions $f^a$ and $f_a$:
$$
f^a=-(\delta^a_1 F_{23} + \delta^a_2F_{31} + \delta^a_3 F_{12}) = (\delta^a_1 C^c_{23} + \delta^a_2 C^c_{31} + \delta^a_3 C^c_{23})\alpha_c,
$$
$$
f_a=-(\delta_a^1 F^{23} + \delta_a^2 F^{31} +\delta_a^3 F^{12}),
$$
Condition \eqref{n20} takes the form:
\begin{equation}\label{n21}
F^{a_1b_1}=\eta^{aa_1}\eta^{bb_1}((\delta_a^2\delta_b^3-\delta_a^3\delta_b^2) F_{23} + (\delta_a^3\delta_b^1-\delta_a^1\delta_b^3) F_{31} + (\delta_a^1\delta_b^2-\delta_a^2\delta_b^1) F_{12}).
\end{equation}
From equations \eqref{n20} \eqref{n21} it follows:
\begin{equation} \label{n22}
f_a = f^b m_{ab},
\end{equation}
where
$$
m_{ab} =\left(\begin{array}{ccc} {\eta^{22}\eta^{33}-\eta^{23}\eta^{23}} & {\eta^{13}\eta^{23}-\eta^{33}\eta^{12}} & {\eta^{12}\eta^{23}-\eta^{22}\eta^{13}}\\ {\eta^{13}\eta^{23}-\eta^{33}\eta^{12}} & {\eta^{11}\eta^{33}-\eta^{13}\eta^{13}} & {\eta^{12}\eta^{13}-\eta^{11}\eta^{23}} \\ {\eta^{12}\eta^{23}-\eta^{22}\eta^{13}} & {\eta^{12}\eta^{13}-\eta^{11}\eta^{23}} & {\eta^{11}\eta^{22}-\eta^{12}\eta^{12}}\end{array}\right).
$$

\begin{enumerate}
\item

Let in the system \eqref {n13} index $i$ equals to zero: $i=0$. Then the equation takes following form:
\begin{equation}\label{n23}
 F^{01}_{,1}- F^{01} \sigma_{,1} = -F_{01,1}+ F_{01} \sigma_{,1} + F^{\quad a}_{1}(\zeta^\alpha_{a,\alpha} - \sigma_{|a}).
\end{equation}
$A_1=\zeta_1^a \alpha(a) = \alpha_1(u^0)$. Then:
$F_{01,1}-F_{01} \sigma_{,1} = \dot{\alpha}_1\sigma_{|1}= \dot{\alpha}_1 C_{1}.$
The equation takes the form:
\begin{equation}\label{n24}
C_1\dot{\alpha}_1 = C^c_{b1}\eta^{ab}\alpha_c C_a.
\end{equation}

\item

Let $i=\alpha$. The set of equations \eqref{n13} takes the form:
\begin{equation}\label{n25}
F^{\alpha i}_{,i}- \sigma_{,i}F^{\alpha i} = F^{\alpha}_{. 1,0} - \dot{\sigma}_{0}F^{\alpha}_{. 1} + F^{\alpha \beta}_{,\beta} - \sigma_{,\beta}F^{\alpha \beta}.
\end{equation}
\end{enumerate}
Since
$$
F^{\alpha\beta} = \delta^\alpha_1 F_0^{ \quad \beta} - \delta^\beta_1 F_0^{ \quad \alpha} + \zeta^\alpha_a\zeta^\beta_bF^{ab},
$$
from \eqref{n25} it follows:
$$
\zeta_\alpha^c(F^{\alpha i}_{,i}- \sigma_{,i}F^{\alpha i})= \delta^c_1 (\dot{F}_{0 1} - \dot{\sigma_0} F_{0 1}) + (\dot{F}^{c}_{\quad 1}- \dot{\sigma}_{0} F^{c}_{\quad 1})+ \delta^c_1(\zeta^\beta_{b,\beta} - \sigma_{|b})F^{\quad b}_{0}
$$
$$
-F^{ \quad a}_{0}\zeta^c_\beta \zeta^\beta_{a,1} + \sigma_{,1}F^{ \quad c}_{0}
 + F^{ab}(\zeta^c_\alpha \zeta^\alpha_{a|b}) + F^{cb}(\zeta^\beta_{b,\beta} - \sigma_{|b})\Rightarrow
$$
\begin{equation}\label{n26}
\exp{\sigma_0}((\delta^c_1\dot{\alpha}_1 + F^{c}_{ \quad 1})\exp{-\sigma_0})_{,0} + \delta^c_1 c_a F^{ \quad a}_{0} - $$
$$C_1F^{ \quad c}_{0} - C^c_{1b}F^{ \quad b}_{0} + C_bF^{cb} + \frac{1}{2}C^c_{ba}F^{ab}=0.
\end{equation}
Let us denote:
\begin{equation}\label{n27}
n^{pq}=\eta^{pq}\exp{(-\sigma_0)} \quad (\det{n^{pq}}=1), \quad y_\alpha= \eta^{1\alpha},
\end{equation}
$$
A=(\dot{\alpha}_1 + y_2 f^3 -y_3 f^2 )\exp{-\sigma_0}, \quad Y = n^{33} f^2 - n^{23} f^3 \quad W = n^{23} f^2 - n^{22} f^3
$$
Then using \eqref{n25} - \eqref{n27} one can obtain the Maxwell equations in the form:
\begin{equation}\label{n28}
\dot{A}+C^{21}(Ay_3 +f^1(y_3 n^{23} - y_2 n^{33}) +y_1 Y) - C^{31}(Ay_2 +f^1(y_3 n^{22} - y_2 n^{23}) +y_1 W)
\end{equation}
$$
+ C^{11}(f^1 \exp{\sigma_0}+ y_3 W - y_2 Y)=0;
$$

\begin{equation}\label{n29}
-\dot{W}+C^{22}(Ay_3 +f^1(y_3 n^{23} - y_2 n^{33}) +y_1 Y +\dot{\alpha}_p n^{p3}) -
\end{equation}
$$
C^{32}(Ay_2 +f^1(y_3 n^{22} - y_2 n^{23}) +y_1 W +\dot{\alpha}_p n^{p2})+ C^{12}(f^1 \exp{\sigma_0}+ y_3 W - y_2 Y)=0;
$$

\begin{equation}\label{n30}
-\dot{Y}+C^{23}(Ay_3 +f^1(y_3 n^{23} - y_2 n^{33}) +y_1 Y +\dot{\alpha}_p n^{p3}) - $$
$$
C^{33}(Ay_2 +f^1(y_3 n^{22} - y_2 n^{23}) +y_1 W +\dot{\alpha}_p n^{p2})
\end{equation}
$$
+ C^{13}(f^1 \exp{\sigma_0}+ y_3 W - y_2 Y)=0;
$$

\begin{equation}\label{n31}
C_1 A-C_2 W - C_3 Y=0.
\end{equation}
It is indicated here:
$$
C^{a1}= C^a_{23}, \quad C^{a2}= C^a_{31}, \quad C^{a3}= C^a_{12}, \quad C_a = C^b_{ba}.
$$

\quad

Throughout the text, the non-holonomic components of the metric tensor $\eta^{\alpha\beta}$ and the electromagnetic potential $\alpha_\alpha$ are called field functions.
The system of Maxwell equations consists of four ordinary differential equations, which include nine field functions. In the solution obtained some of these functions will be independent arbitrary functions of the variable $u^0$. The remaining field functions (dependent functions) must be expressed using Maxwell equations in terms of independent functions. The set of solutions of Maxwell equations is divided into two subsets. In the first set solutions contain dependent field functions expressed only in terms of independent field functions and their derivatives with respect to $u^0$. The second set includes solutions in which the dependent field functions are expressed through a system of independent functions and their derivatives with respect to the variable $u^0$ (some of the functions included in this system are not field functions).

In the first case, Maxwell equations represent a non-degenerate system of linear algebraic equations for the dependent field functions. The coefficients of these functions on the left-hand sides of the equations in this system, and the right-hand sides of the equations, contain only the independent field functions and their derivatives with respect to the variable $u^0$.

In the second case, such a non-degenerate system does not exist. To find solutions of Maxwell equations, it is necessary to find a new system of independent functions through which the dependent field functions  can be expressed.

In the present paper, the classification problem is solved for all invariant electromagnetic fields in homogeneous Petrov null-spaces with solvable groups of motions $G_3(N) (N\leq VIII)$. All appropriate non-equivalent exact solutions of vacuum Maxwell equations are found.

\section{\quad Solutions of Maxwell equations for solvable  groups \quad ${\bf G_3(N\leq VII)}$}

When obtaining the Killing vector fields of the null homogeneous  Petrov space of type $V^{*}_{4} (N)$ from the Killing fields of the invariant space $V_3(N)$, one can use only those admissible transformations of the variables [$u^\alpha$], which do not destroy the condition:
\begin{equation}\label{n32}
\xi^\alpha_{a,1}=0.
\end{equation}
From this condition it follows that to each nonequivalent set of Killing vector fields of the motions group $G_{3}(N)$ which acts on the invariant space $V_{3}(N)$ there correspond $1 \leq n\leq3$ nonequivalent sets of Killing vector fields of the motions groups $G_{3}(N)$ of the space $V^{*}_{4}(N)$. The number $n$ has the following values ( see \cite{3n}-\cite{5n}):

\begin{enumerate}

\item For the spaces $G_3(I), G_3(IX), n=1; $

\item For the spaces  $G_3(II), G_3(VIII), n=2; $

\item For the spaces $G_3(III) - G_3(VIII), n=3. $

\end{enumerate}

The following subsections of the paper present all nonequivalent solutions of Maxwell vacuum equations for all homogeneous null Petrov spaces of type $V^*_4(N_{n})$ with solvable motions groups.

The homogeneous null Petrov spaces with non-solvable groups of types $G_3(VIII)$ and $G_4(VIII)$ will be considered in another paper.

\subsection{\textbf{Spaces of type} $V^*_4(II_{n})$.}

There are only two nonequivalent null homogeneous Petrov spaces of type $V^*_4(II_{n})$. Furthermore, for each motions group $G^*_3(N_{n})$ of spaces $V^*_4(N_{n})$, complete sets of Killing vector fields $\xi^\alpha_a$ are presented. They are used to find the vector fields of the canonical frame $\zeta^\alpha_a$ from the system of equations \eqref{n5}. Thus, the sets of vector fields $\xi^\alpha_a$ and $\zeta^\alpha_a$ correspond to each other. At the same time the sets of structural constants $\tilde{C}^a_{bc} \quad ([X_a,X_b]=\tilde{C}^c_{ab}X_c)$ and $C^a_{ab} \quad ([Y_a,Y_b]=C^c_{ab}Y_c)$ may not coincide.

In the space $V^*_4(N_{n})$, whose metric tensor satisfies the equations \eqref{n5}, the Hamilton-Jacobi equation admits a complete set of integrals of motion corresponding to a complete set of Killing vector fields of the group $G^*_3(N_{n})$

\subsubsection{\textbf{Spaces of type} $V^*_4(II_{1})$.}

The group operators are:
$$
X_a= \xi^\alpha_a p_a:\quad X_a= \delta_a^1 p_1 + \delta_a^2 p_2 + \delta_a^3 (u^2 p_1 + p_3),
\quad Y_a= \zeta^\alpha_a p_a = X_a. 
$$
There is only one non-zero structural constant \quad $C^1_{23}= C^{11}=1.$\quad Hence:
$$
\quad f^1 =\alpha_1, \quad f^2 =f^3 =0.
$$
Maxwell equations \eqref{n28}-\eqref{n31} reduce to the system of two equation:
$$
\dot{A} + \alpha_1 \exp{\sigma_0} =0, \quad \dot{\alpha}_1 = A\exp{\sigma_0} \rightarrow \sigma_0 = \ln{\dot{\beta}}, \quad \alpha_1 = e \sin{\beta} \quad A = e \cos{\beta}.
$$
The solution of the Maxwell equations has the form:
$$
\alpha_1 = e \sin{\beta}, \quad \eta^{22} = \frac{(\dot{\beta})^2 + (\eta^{23})^2}{\eta^{33}}, \quad(e =const).
$$
The functions $\eta^{22}\quad \eta^{33}$ are the depended field functions. The function $\beta$ and all other field functions are the independent functions.

\subsubsection{\textbf{Spaces of type} $V^*_4(II_{2})$.}

The group operators have the following form:
$$
X_a= \xi^\alpha_a p_a:\quad X_a= \delta_a^1 p_2 + \delta_a^2 p_3 + \delta_a^3 (p_1 + u^3 p_2 ),
$$
$$
Y_a= \zeta^\alpha_a p_a = \delta_a^1 p_1 + \delta_a^2 (p_2 + u^1 p_3) + \delta_a^3 p_3.
$$
There is only one nonzero structural constant \quad $C^3_{12} =C^{33}=1.$ \quad Hence:
$$
\quad f^1 =f^2=0,\quad f^3 = \alpha_3.
$$
Maxwell equations \eqref{n28}-\eqref{n31} reduce to equations:
$$
\dot{A} =0 \Rightarrow A = a, \quad \dot{W} =0 \Rightarrow W = -b
\Rightarrow \alpha_3 =\frac{b}{n^{22}} \quad  (a, b = const).
$$
\begin{equation}\label{n33}
\dot{Y} + a y_2 + \alpha_p n^{p2} + y_1 b =0.
\end{equation}
Depending on whether the constant $b$ is equal to zero or not, there are two possible solutions.

\quad

\begin{enumerate}

\item
$b=0 \Rightarrow ay_2 + \dot{\alpha}_2 n^{22}$=0. Electromagnetic field disappears if {a=0}. Therefore $a=\frac{1}{e}\ne 0  \Rightarrow$
$$
 y_2=-e\dot{\alpha}_2 n^{22}, \quad \sigma_0 = \ln(e\dot{\alpha}_1).
$$
The solution is:
$$
\alpha_3 =0, \quad \eta^{12}=-\frac{\dot{\alpha}_2}{\dot{\alpha}_1}\eta^{22}, \quad \eta^{33} = \frac{(e\dot{\alpha}_1)^2 + (\eta^{23})^2}{\eta^{22}}.
$$
$\eta^{12}\quad \eta^{33}$ are the depended field functions. All other field functions are independent functions.

\quad

\item
$b\ne 0.$ From equations \eqref{n33} it follows the solution:
$$
\quad \eta^{12} = \frac{n^{22}}{b}(a \exp{\sigma_0} - \dot{\alpha}_1), \quad \eta^{11} = -\frac{\dot{n^{23}}}{n^{22}} + \frac{an^{22}}{b^2}(a \exp{\sigma_0} - \dot{\alpha}_1) + \frac{\dot{\alpha}_2}{b},$$
$$ \eta^{33}=\frac{(\eta^{23})^2 + \exp{2\sigma_0} }{\eta^{22}},\quad \alpha_3 = \frac{b}{n^{22}}.
$$
Functions $\alpha_3, \quad \eta^{11}\quad \eta^{12}, \quad \eta^{33} $ are the depended field functions $(n^{pq}=\eta^{pq}\exp{-\sigma_0})$. All other field functions  are the independent functions.

\end{enumerate}

\subsection{ \textbf{Spaces} \quad $V^*_4(III_{n})$}

\subsubsection{\textbf{Spaces of type }$V^*_4(III_{1})$}.

The operators $X_a, Y_a$ have the form:
$$
X_a= \delta_a^1 p_1 + \delta_a^2 p_2 + \delta_a^3 (u^2 p_2 + p_3), \quad
Y_a= \delta_a^1 p_1 + \delta_a^2 p_2\exp{-u^3} + \delta_a^3 p_3.
$$
There is one non-zero structure constant \quad $C^2_{23} = C^{21}=1.$\quad Hence:

$$
f^1 = \alpha_2, \quad f^2 =f^3 =0,\quad C_3 =1 \Rightarrow W=Y=0.
$$
Maxwell equations \eqref{n28} \eqref{n31} are as follows:
$$
A = 0 \Rightarrow \dot{\alpha}_1 = c = const, \quad \alpha_2(y_3n^{23}- y_2 n^{33}) = 0, \quad \dot{\alpha}_1 = A\exp{\sigma_0} =0.
$$
The solution of this system has the form:

\begin{enumerate}

\item $\alpha_2 = 0, \quad \alpha_1 = c = const, $ \quad Functions $\quad \alpha_3, \eta^{ab}$ \quad are independent field functions.

\quad

\item $\alpha_1 = 0,$
$$
\eta^{12}= \frac{\eta^{13}\eta^{23}}{\eta^{33}},\quad
$$
Functions $\eta^{12}$ is the depended field functions,  functions $\eta^{ab} \alpha_{p}$ and all other field functions are the independent functions.

\end{enumerate}

\quad

\subsubsection{\textbf{Spaces of type $V^*_4(III_{2})$}}

\quad

The operators $X_a, Y_a$ have the form:
$$
X_a= \delta_a^1 p_2 + \delta_a^2 p_3 + \delta_a^3 (p_1 + u^3 p_3 ),
\quad Y_a= \delta_a^1 p_1 + \delta_a^2 p_2 + \delta_a^3 p_3\exp{u^1}.
$$
There is one non-zero structure constant \quad $C^3_{13} = -C^{32}=1.$\quad Hence: $$ C_1 = -1 \quad f^1 =f^3 =0,\quad f^2 = -\alpha_3.$$
Maxwell equations \eqref{n28} \eqref{n31} are of the form:
\begin{equation}\label{n34}
A =0 \Rightarrow \dot{\alpha}_1 +y_3\alpha_3 =0, \quad \dot{Y} =0 \Rightarrow n^{23}\alpha_3 = e, \quad W = n^{22}\alpha_3 \quad (e = const).
\end{equation}
From \eqref{n34} it follows:
$$
\alpha_3 =\frac{e}{n^{33}}\quad (e=const), \quad \dot{\alpha_2}n^{22}n^{33}=e(\dot{n^{23}} +y_1n^{23}).
$$
In the case $\alpha_3 = 0$ the electromagnetic field disappears. Therefore, we assume that $\alpha_3 \ne 0 \rightarrow y_3 = -\frac{n^{33}\dot{\alpha_1}}{e}.$
The solutions of the system of equations \eqref{n34} can be represented as:

\begin{enumerate}
  \item $\eta^{23}\ne 0.$
$$
\eta^{11}=\dot{\sigma_0} - (\ln{\eta^{23}})_{,0} + \frac{\dot{\alpha}_2 }{e\eta^{23}\exp{\sigma_0}}((\eta^{23})^2 + \exp{2\sigma_0}), \quad $$
$$\eta^{22} = \frac{(\eta^{23})^2 + \exp{2\sigma_0}}{\eta^{33}}, \quad \eta^{13} = -\frac{\eta^{33}\dot{\alpha_1}}{e\exp{\sigma_0}}, \quad
\alpha_3=\frac{e \exp{\sigma_0}}{\eta^{33}}.
$$
The functions \quad $\alpha_3, \quad \eta^{11}\quad \eta^{13}, \quad \eta^{22}\quad $ are the depended field functions. All other field functions  are the independent functions.

\quad

  \item $\eta^{23}=0$,
  $$
   \eta^{13} = -\frac{\eta^{33}\dot{\alpha_1}}{e\exp{\sigma_0}}, \quad \eta^{22} = \frac{\exp{2\sigma_0}}{\eta^{33}},\quad \alpha_3=\frac{e \exp{\sigma_0}}{\eta^{33}}.
   $$
The functions $\alpha_3, \quad \eta^{11}, \quad \eta^{22}\quad $ are the depended field functions. All other field functions  are independent functions.
\end{enumerate}

\subsubsection{\textbf{Spaces of type }$V^*_4(III_{3})$}

The operators $X_a, Y_a$ have the form:
$$
X_a= \delta_a^3 p_3 + \delta_a^2 p_2 + \delta_a^1 p_1\exp{u^3},
\quad Y_a= \delta_a^1 p_1 + \delta_a^2 p_2 + \delta_a^3 (p_3-u^1p_1).
$$
There is one non-zero structure constant \quad $C^1_{13} = -C^{32}=-1.$\quad Hence:
$$ C_3 = -1 \quad f^1 =f^3 =0,\quad f^2 = \alpha_1.$$
Maxwell equations \eqref{n28} \eqref{n31} have the form:
\begin{equation}\label{n35}
Y =0 \Rightarrow f^2 = \alpha_1 =0 \Rightarrow W=0, \quad A =0.
\end{equation}
The single solution is:
$$
\alpha_1=0.
$$
All other field functions  are independent functions.

\subsection{ \textbf{Spaces} \quad $V^*_4(IV_{n})$}

\subsubsection{ \textbf{Spaces} \quad $V^*_4(IV_{1})$}

The operators $X_a, Y_a$ have the form:
$$
X_a= \delta_a^1 p_1\exp{-u^3} + \delta_a^2 p_2 + \delta_a^3( u^2(p_2 + p_1\exp{-u^3})+p_3), $$
$$
Y_a= \delta_a^1 p_1 + \delta_a^2(u^3p_1 + p_2\exp{u^3}) + \delta_a^3 (u^1 p_1 - p_3).
$$
Non-zero structure constants are: \quad $C^1_{13}=C^1_{23} =C^2_{23} = 1\Rightarrow $
$$\quad
C^{12}=-1, \quad C^{11}=C^{21}=1, \quad f^1 = \alpha_1 + \alpha_2,\quad f^2 = -\alpha_1, \quad f^3 =0,\quad C_a = 2\delta_a^3.
$$
Maxwell equations \eqref{n28} \eqref{n31} have form:
$$
Y = f^2 n^{33} = 0 \Rightarrow f^2 = \alpha_1 = 0 , \quad W =0 \Rightarrow \quad f^1 = \alpha_2 = A=0.
$$
This leads to a unique solution:
$$
\alpha_1 = \alpha_2 =0.
$$
The remaining field functions $(\alpha_3, \eta^{ab})$ are independent functions.

\subsubsection{ \textbf{Spaces} \quad $V^*_4(IV_{2})$}

The operators $X_a, Y_a$ have the form:
$$
X_a= \delta_a^1 p_2 + \delta_a^2 p_3 + \delta_a^3 (p_1 + (u^2 +u^3) p_2 + u^3p_3), $$
$$
Y_a= \delta_a^1 p_1 + \delta_a^2 p_2u^1 + \delta_a^3 (u^1 p_2 + p_3)\exp{u^1}.
$$
Non-zero structure constants: \quad $C^2_{12}=C^2_{13} =C^3_{13} = 1 ,\quad C_a = -2\delta_a^1 \Rightarrow $
\begin{equation}\label{n36}
C^{23}=1, \quad C^{22}=C^{32}=-1, \quad f^1 =0,\quad f^2 =-(\alpha_2 + \alpha_3), \quad f^3 =\alpha_2.
\end{equation}
Maxwell equations \eqref{n28} \eqref{n31} have form:

\begin{equation}\label{n37}
A=0 \Rightarrow \dot{\alpha}_1+ y_2f^3 -y_3 f^2 =0,
\end{equation}
\begin{equation}\label{n38}
\dot{Y}=\dot{\alpha}_p n^{p3} +y_1 Y,
\end{equation}
\begin{equation}\label{n39}
\dot{Y}+ \dot{W}=\dot{\alpha}_p n^{p2} +y_1 W,
\end{equation}
We eliminate the function $y_1$ from the system of equations \eqref{n38}, \eqref{n39}. Then we get:
\begin{equation}\label{n40}
Y(\dot{Y}+ \dot{W}) - \dot{Y}W = (\dot{\alpha}_3\alpha_2 - \dot{\alpha}_2\alpha_3 - \dot{\alpha}_2\alpha_2)
\end{equation}
(the relation used is: $\dot{\alpha}_p (n^{p2}Y - n^{p3}W)= \dot{\alpha}_p f^p$.
When  ${f^2}^2+{f^3}^2 \ne 0.$ the functions $Y, W $ can be considered as independent functions, and the
functions $n^{ab}$ can be expressed in terms of them and the functions $f^p$ as follows:
\begin{equation}\label{n41}
n^{22}=\frac{(f^2)^2 + W^2}{f^2 Y - f^3 W}, \quad n^{23}=\frac{f^2f^3 + WY}{f^2 Y - f^3 W}, \quad n^{33}=\frac{(f^3)^2 + Y^2}{f^2 Y - f^3 W}.
\end{equation}
Using \eqref{n36} one can represent \eqref{n41} in the form:
\begin{equation}\label{n42}
n^{22}=-\frac{(\alpha_2+ \alpha_3)^2 + W^2}{(\alpha_2+ \alpha_3) Y + \alpha_2 W}, \quad $$
$$
n^{23}=\frac{\alpha_2(\alpha_2+ \alpha_3) - WY}{(\alpha_2+ \alpha_3) Y + \alpha_2 W}, \quad n^{33}=-\frac{(\alpha_2)^2 + Y^2}{(\alpha_2+ \alpha_3) Y + \alpha_2 W};
\end{equation}
Using \eqref{n38}, \eqref{n39} - \eqref{n42},  one can find the function $y_1$.

\begin{enumerate}

\item $Y \ne 0 \rightarrow$
\begin{equation}\label{n43}
\eta^{11} = \frac{\dot{Y}}{Y}-\frac{(\dot{\alpha}_2(\alpha_2(\alpha_2 + \alpha_3)) - YW)-\dot{\alpha}_3((\alpha_2)^2 + Y^2)}{Y((\alpha_2+ \alpha_3) Y + \alpha_2 W)}
\end{equation}
$$(y_1 =\eta^{11} = \frac{\dot{Y}}{Y}-\frac{\dot{\alpha}_p n^{p3}}{Y}).$$

\item $Y=0 \rightarrow $

\begin{equation}\label{n44}
\eta^{11} = \frac{\dot{Y}}{Y}-\frac{(\dot{\alpha}_3(\alpha_2(\alpha_2 + \alpha_3)) - YW)-\dot{\alpha}_2((\alpha_2+\alpha_3)^2 + W^2)}{Y((\alpha_2+ \alpha_3) Y + \alpha_2 W)}
\end{equation}
$$ ( y_1 = \eta^{11} = \frac{\dot{W}}{W}-\frac{\dot{\alpha}_p n^{p2}}{W} ).$$
\end{enumerate}
The solutions of equations \eqref{n40}, \eqref{n37} are.

\begin{enumerate}

\item $f^3 = \alpha_2 \ne 0 \quad$
 $$
y_2=\frac{\dot{\alpha}_1+y_3f^2}{\alpha_2} \Rightarrow \eta^{12}=\frac{\dot{\alpha}_1-\eta^{13}(\alpha_2 + \alpha_3))}{\alpha_2}.
 $$

In this case, equation \eqref{n37} has solutions:

\begin{enumerate}

\item
\begin{equation}\label{n45}
(\alpha_2)^2 = \frac{1}{\dot{\omega}}(\dot{Y}Y + Y\dot{W} - W\dot{Y}),
\quad \alpha_3 =\alpha_2(\omega +\ln\alpha_2),
\end{equation}
The functions \quad $ \eta^{13},\quad\eta^{11}, \quad  \alpha_1,\quad \sigma_0, \quad \eta^{13},\quad  Y,\quad  W ,\quad \omega $ \quad are independent functions.

\item

\begin{equation}\label{n46}
\alpha_3 =\alpha_2\ln\alpha_2, \quad W=-Y\ln{aY} \quad (a= const).
\end{equation}
Functions
$$\alpha_2,\quad \alpha_1,\quad  \eta^{13},\quad  \sigma_0,\quad  \omega, \quad Y,\quad  W $$
are the independent functions.

The functions \quad $ \eta^{13},\quad\eta^{11}, \quad  \alpha_1,\quad  \eta^{13},\quad  Y,$ \quad are independent functions.

\item $Y=0$

\begin{equation}\label{n47}
\alpha_3 =\alpha_2\ln\alpha_2, \quad \eta^{11}= y_1 = \frac{\dot{W}}{W}-\frac{\dot{\alpha}_p n^{p2}}{W}
\quad (see \quad (42)). \end{equation}

The functions.
$$\alpha_1, \quad \alpha_2, \quad \eta^{13},\quad  \sigma_0,\quad  W \quad $$
are independent functions.

\quad

$\eta^{pq}=n^{pq}\exp\sigma_0 \quad$ (see \quad \eqref{n42}).

\quad

\end{enumerate}

\item

$\alpha_2 =0$. The solution is:
$$
W=-Y\ln{qY}, \quad y_3 =\eta^{13} =-\frac{\dot{\alpha}_1}{\alpha_3}.
$$

The functions
$$\alpha_1, \quad \alpha_3, \quad \eta^{12},\quad  \sigma_0,\quad  Y\quad $$
are independent functions.

\quad

$\eta^{pq}=n^{pq}\exp\sigma_0$ \quad (see \quad \eqref{n42},

\quad

\end{enumerate}
\quad

\subsubsection{ \textbf{Spaces} \quad $V^*_4(IV_{3})$}

Operators $X_a, Y_a$ have the form:
$$
X_a= \delta_a^1 p_1\exp{-u^3} + \delta_a^2 p_2 + \delta_a^3 (p_1\exp{-u^3}+p_2 + p_3), $$
$$
Y_a= \delta_a^1 p_1 + \delta_a^2 p_2\exp{u^3} + \delta_a^3(p_1u^1 +u^1 \exp{u^3}p_2 - p_3).
$$
Non-zero structure constants: \quad $C^1_{13}=C^2_{13} =C^2_{23} = 1 ,\quad c_a = 2\delta_a^3 \Rightarrow $
$$\quad
C^{12}= C^{22}=-C^{21}=-1, \quad f^1 =\alpha_2,\quad f^2 =-(\alpha_1 + \alpha_2), \quad f^3 =0.
$$
The Maxwell equations \eqref{n28} \eqref{n31} have the form:

\begin{equation}\label{n48}
Y=0 \Rightarrow f^2 =0,\quad \alpha_2 = -\alpha_1, \quad W=0 \Rightarrow \dot{\alpha}_1= A\exp\sigma_0,
\end{equation}
\begin{equation}\label{n49}
\dot{A} +A y_3 =\alpha_1(n^{23}y_3 - n^{33}y_2),
\end{equation}
\begin{equation}\label{n50}
\dot{A} +\alpha_1 \dot{\alpha}_1 \frac{1}{A} =-\dot{\alpha}_1n^{23} + \dot{\alpha}_3n^{33},
\end{equation}
Obviously, $\alpha_1 \ne 0$ (otherwise the electromagnetic field disappears). Depending on whether the determinant of the system of algebraic equations \eqref{n49}-\eqref{n50} $\Delta=\alpha_1(\dot{\alpha}_3 y_3 - \dot{\alpha}_1 y_2) $ is equal to zero or not, the following solutions exist.

\begin{enumerate}

\item
$$
\alpha_1 =- \alpha_2 = \frac{1}{e} =const, \quad \eta^{23}=\frac{1}{e}(\dot{\alpha}_3\eta^{33}-e\exp2\sigma_0),.
$$
$$
\eta^{12} = \frac{\eta^{13}}{e}(\dot{\alpha}_3\eta^{33}\exp(-\sigma_0)-e\exp\sigma_0), \quad
\eta^{22}= \frac{1}{\eta^{33}}\left((\eta^{23})^2 + \exp{2\sigma_0}\right).
$$
The functions,\quad $\alpha_{3}, \quad \eta^{11},\quad \eta^{33}, \quad \eta^{13}, \quad \sigma_0 \quad $ are independent field functions.

\item

$$ \alpha_1 =- \alpha_2,\quad
\eta^{13} = - \frac{A\dot{A}\dot{\alpha}_1}{A^2 \dot{\alpha}_1 +\alpha_1 A\dot{A} +\dot{\alpha}_1\alpha_1^2},\quad \eta^{12}= - \frac{A\dot{A}\dot{\alpha}_3}{A^2 \dot{\alpha}_1 +\alpha_1 A\dot{A} +\dot{\alpha}_1\alpha_1^2}, \quad
$$
$$
\eta^{23}= \left(A\dot{\alpha}_3\eta^{33} -\frac{{\dot{\alpha}_1}}{A}(A\dot{A} +\dot{\alpha}_1\alpha_1)\right)\frac{1}{\dot{\alpha}_1 A},\quad \eta^{22}= \frac{1}{\eta^{33}}\left((\eta^{23})^2 + \left(\frac{{\dot{\alpha}_1}}{A}\right)^2\right).
$$
The functions $\eta^{12},\quad \eta^{13}, \quad \eta^{23}, \quad \eta^{22} \quad$ are depended field functions,\quad

The functions $\alpha_{1},\quad A \quad \alpha_{3},\quad \eta^{11},\quad $ are independent functions.

\item $\quad \alpha_1 =- \alpha_2 $

$$
\eta^{23} = \frac{{\dot{\alpha}_1}}{A(\dot{\alpha}_3 \eta^{13} - \dot{\alpha}_1 \eta^{12})}\left(\eta^{12}\frac{A\dot{A} +\dot{\alpha}_1\alpha_1}{A} + \dot{\alpha}_3\frac{\dot{A}+\eta^{13} A}{\alpha_1}\right),\quad
$$
$$
\eta^{22}= \frac{1}{\eta^{33}}\left((\eta^{23})^2 + \left(\frac{{\dot{\alpha}_1}}{A}\right)^2\right), $$
$$
\eta^{33} = \frac{{\dot{\alpha}_1}}{A(\dot{\alpha}_3\eta^{13} - \dot{\alpha}_1 \eta^{12})}\left(\eta^{13}\frac{A\dot{A} +\dot{\alpha}_1\alpha_1}{A} + \dot{\alpha}_1\frac{\dot{A}+\eta^{13} A}{\alpha_1}\right).
$$
The functions,\quad $\alpha_{1},\quad \alpha_{3},\quad A\quad \eta^{1\alpha},\quad $ are independent functions.
\end{enumerate}

\subsection{ \textbf{Spaces \quad $V^*_4(V_{n}), \quad V^*_4(VI_{n})$}}

The group of motions of a space of type $V_3(VI)$ is defined by the structure constants $C^a_{bc}$, which can be reduced to canonical form by choosing the group operators $X_a$:
$$
C^a_{bc} = q\delta^a_2(\delta^2_b\delta^3_c -\delta^2_c\delta^3_b) + \delta^a_1(\delta^1_b\delta^3_c -\delta^1_c\delta^3_b).
$$
$q$ is an arbitrary parameter. When $q=1$, the group $G^*_4(VI_{n})$ becomes the group $G^*_4(V_{n})$.
(See Petrov\cite{3n}, \cite{4n}). Therefore the group $G^*_4(V_{n})$ will not be considered separately.

\subsubsection{\textbf{Spaces of type} $\quad V^*_4(VI_{1})$}.

Operators $X_a, Y_a$ have the form:
$$
X_a= \delta_a^1 p_1\exp{(-u^3)} + \delta_a^2 p_2 + \delta_a^3( q u^2 p_2+p_3), $$
$$
Y_a= \delta_a^1 p_1 + \delta_a^2 p_2\exp{qu^3} + \delta_a^3 (u^1 p_1 - p_3).
$$
Non-zero structure constants: \quad $C^1_{13}=1 \quad C^2_{23} = q \Rightarrow $
$$\quad
C^{12}=-1, \quad C^{21}=q, \quad f^1 = q\alpha_2,\quad f^2 = -\alpha_1, \quad f^3 =0,\quad C_a = \delta_a^3(1+q).
$$
From \eqref{n31} it follows: $Y(1+q)= 0.$ Therefore, it is necessary to consider the following cases.

\begin{enumerate}

\item

$Y =0.$ 

The Maxwell equations \eqref{n28} \eqref{n31} are:
$$
Y = 0 \Rightarrow f^2 = \alpha_1 = 0 , \quad W =0 \Rightarrow \quad f^1 = \alpha_2 = A=0.
$$
From this it the solution follows:
$$
\alpha_1 = \alpha_2 =0.
$$
The remaining field functions $(\alpha_3, \eta^{ab})$ are independent functions.

\quad

\item
$q=-1 \quad (Y\ne 0).$

The Maxwell equations \eqref{n28} - \eqref{n30} have the form:
$$
\dot{A}-Ay_3 +\alpha_2(y_3n^{23} -y_2 n^{33}) +e y_1 =0, \quad A \exp\sigma_0 =\dot{\alpha_1} + \alpha_1 y_3, \quad\dot{Y}=0 \Rightarrow
$$
$$
(n^{23}\alpha_1)_{,0} +y_3(n^{23}\alpha_1) +\alpha_2\exp\sigma_0 = e y_2.
$$
From this follows the solution:
$$
n^{33} = \frac{e}{\alpha_1},\quad n^{22} = \frac{\alpha_1}{e}\left(1 + (n^{23})^2\right),\quad \eta^{1\alpha}=y_\alpha
$$
$$
\eta^{11}=\frac{e}{\alpha_1}\left((A^2 + \alpha_2^2)\exp\sigma_0-(A\alpha_1)_{,0} +(n^{23} \alpha_1)_{,0}\alpha_2\right),
$$
$$
\eta^{12}= e((n^{23}\alpha_1)_{,0} + (An^{23} + \alpha_2)\exp\sigma_0 - n^{23} \dot{\alpha}_1), \quad \eta^{13}=\frac{1}{\alpha_1}(A\exp\sigma_0 -\dot{\alpha}_1 ).
$$
The functions \quad $\eta^{23}, \quad \alpha_{1},\quad \alpha_{3},\quad \sigma_0\quad A, \quad $ are independent field functions.
\end{enumerate}

\subsubsection{\textbf{Spaces of type} $\quad V^*_4(VI_{2})$}.

Operators $X_a, Y_a$ have the form:
$$
X_a= \delta_a^1 p_2 + \delta_a^2 p_3+ \delta_a^3(p_1 +u^2 p_2+ q u^3 p_3);$$
$$
Y_a= \delta_a^1 p_1 + \delta_a^2 p_2\exp{u^1} + \delta_a^3 p_3\exp{qu^1}.
$$

Non-zero structure constants are: $\quad C^2_{12}= 1, \quad C^3_{13}=q \quad \Rightarrow$
$$
C^{13}=1,\quad C^{32}=-q, \quad C_a = -\delta_a^1(1+q),\quad   f^1 =0, \quad f^2 = -q\alpha_3,\quad f^3 = \alpha_2.
$$
From \eqref{n31} it follows: $A(1+q)= 0.$ Therefore, it is necessary to consider the following cases.

\begin{enumerate}

\item

$A=0$

Maxwell equations \eqref{n28} \eqref{n31} have the form:
\begin{equation}\label{n51}
\quad \dot{Y} = \dot{\alpha}_pn^{p3} + y_1 Y, \quad \dot{W} = q(\dot{\alpha}_pn^{p2} + y_1 W).
\end{equation}
$$
\dot{\alpha_1}+ y_2 \alpha_2 + y_3q\alpha_3 = 0.
$$
Obviously, $Y^2 + W^2 \ne 0$. Otherwise, the electromagnetic field disappears. Therefore, the function $y_1$ can be eliminated from the system of equations \eqref{n48}. As a result, we obtain equation:
\begin{equation}\label{n52}
\dot{W}Y -qW\dot{Y} =q(\alpha_2\dot{\alpha}_3 -q\dot{\alpha}_2\alpha_3).
\end{equation}
Let's consider the possible options.

\quad

\begin{enumerate}

\item $\alpha_2 =0$.

\quad

Obviously, $Y\ne0$ otherwise the field disappears. From \eqref{n49} it follows:

$\quad W= c Y^q. \quad$ Using equations  \eqref{n48},  \eqref{n41} one can obtain the solution:
$$
\eta^{22}=-\frac{(q\alpha_3)^2 +(cY^q)^2}{q\alpha_3 Y}\exp{\sigma_0}, \quad \eta^{23}=-\frac{cY^q}{q\alpha_3 }\exp{\sigma_0},
$$

$$\quad \eta^{33}=-\frac{Y}{q\alpha_3}\exp{\sigma_0}, \quad c=const.
$$
The functions \quad $\alpha_1, \quad \alpha_3, \quad \sigma_0,\quad Y$ \quad $\eta^{12}$, \quad are independent.

\quad

\item $\alpha_2 \ne 0.$

\quad

 The function $\alpha_3$ has form:  $\alpha_3 =\omega(u^0) (\alpha_2)^q.$ Substitute this into equation \eqref{n49}. As a result, we find the solution of the Maxwell equations:
$$
\alpha_2 =\left(\frac{\dot{W}Y -qW\dot{Y}}{q\dot{\omega}}\right)^{\varsigma}, \quad \alpha_3 = \omega\left(\frac{\dot{W}Y -qW\dot{Y}}{q\dot{\omega}}\right)^{q\varsigma} \quad \left(\varsigma= \frac{1}{q+1}\right).
$$
$$
\eta^{22}=-\frac{(q\alpha_2)^2 + W^2}{q\alpha_3 Y + \alpha_2 W}\exp{\sigma_0}, \quad \eta^{23}=\frac{WY-q\alpha_2\alpha_3 }{q\alpha_3 Y + \alpha_2 W}\exp{\sigma_0},
$$
$$
\eta^{33}=-\frac{(\alpha_2)^2 + Y^2}{q\alpha_3 Y + \alpha_2 W}\exp{\sigma_0}, \quad\eta^{11}= \frac{\dot{Y}}{Y} -\frac{\dot{\alpha}_p \eta^{p3}\exp{-\sigma_0}}{Y}, \quad
$$
$$
\eta^{12} =-\frac{(\dot{\alpha}_1+q\alpha_3 \eta^{13})}{\alpha_2}, \quad
\eta^{11}= y_1, \quad \quad \eta^{12}= y_2. $$
The field functions \quad $\eta^{13} \quad \alpha_1, \quad \alpha_3, \quad \sigma_0,\quad Y\quad W \quad $, are independent functions.

\end{enumerate}

\quad

\item $q=-1, \quad A = (e)^{-1} = const.$

\quad

The Maxwell equation takes the form: \quad $\exp{\sigma_0} = e(\dot{\alpha}_1 +y_2\alpha_2 -y_3 \alpha_3)$,
$$
(\alpha_3n^{23}-\alpha_2n^{22})_{,0} = -\left(\frac{y_2}{e} +  \dot{\alpha_p}n^{2p} +y_1(\alpha_3n^{23}-\alpha_2n^{22})\right),
$$
$$
 (\alpha_3n^{33}-\alpha_2n^{23})_{,0} = \left(\frac{y_3}{e} +  \dot{\alpha_p}n^{3p} +y_1(\alpha_3n^{33}-\alpha_2n^{23})\right).
$$
The solution may be represented in the form:\quad
$\eta^{pq}=n^{pq}(\dot{\alpha}_1 +\eta^{12}\alpha_2 - \eta^{13}\alpha_3), $
$$
\eta^{12} =e(\alpha_3n^{23}-\alpha_2n^{22})_{,0}-n^{23}(\dot{\alpha_2} -\eta^{11}\alpha_2) -n^{33}(\dot{\alpha_3} +\eta^{11}\alpha_3), \quad n^{22}=\frac{(n^{23})^2 + 1}{n^{33}},
$$
$$
\eta^{13} =e(\alpha_3n^{33}-\alpha_2n^{23})_{,0}-n^{22}(\dot{\alpha_2} -\eta^{11}\alpha_2) -n^{23}(\dot{\alpha_3} +\eta^{11}\alpha_3).
$$
Functions $\alpha_\beta, n^{22}, n^{33}$ are independent field functions.

\end{enumerate}

\subsubsection{\textbf{Spaces of type }$V^*_{4}({VI}_3)$}

Operators $X_a, Y_a$ have the form:
\begin{equation} \label{n53}
X_{1} =p_{1}\exp(-qu^{3}) + p_{2}\exp((1-q)u^{3}), \quad X_{2} =p_{2}, \quad X_{3} =p_{3} - p_2\exp(u^{3}).
\end{equation}
\begin{equation} \label{n54}
{\rm Y}_{1} =p_{1}, \quad Y_{2} =p_{2} \exp{u^3}, \quad Y_{3} =p_{3}-u^1(q p_{1} +p_2(q-1)\exp{u^3}).
\end{equation}
Non-zero structure constants are: \quad $ C^1_{13}= -q, \quad C^2_{13}=(1-q), \quad C^2_{23}=-1 \quad \Rightarrow$
$$ C_a = -\delta_a^3(1+q),\quad C^{12}=q, \quad C^{22}=q-1, \quad C^{21}=-1,\Rightarrow
$$
$$
f^1 =-\alpha_2,\quad f^2 = q\alpha_1 +(q-1)\alpha_2,\quad f^3 = 0.
$$
Maxwell equations \eqref{n28} \eqref{n31} have the form:
\begin{equation}\label{n55}
Y(q+1)=0,\quad \dot{Y} = 0 \Rightarrow Y = \tilde{e} =const \quad ((q+1)\tilde{e} =0)
\Rightarrow
\end{equation}
$$
f^2 =\frac{\tilde{e}}{n^{33}}, \quad W =\frac{\tilde{e} n^{23}}{n^{33}},
\quad A = (\dot{\alpha}_1 - \frac{\tilde{e} y_3}{n^{33}})\exp{-\sigma_0}.
$$
\begin{equation}\label{n56}
\left(\frac{\tilde{e} n^{23}}{n^{33}}\right)_{,0} = (q-1)(\dot{A} + \dot{\alpha}_p n^{p3}) +
q(\tilde{e}(y_3\frac{n^{23}}{n^{33}}-y_2)-\alpha_2\exp{\sigma_0}),
\end{equation}
$\dot{A} = y_3(A -\alpha_2 n^{23}) + y_2\alpha_2 n^{33} + \tilde{e} y_1.$

\begin{enumerate}

\item $ q+1 \ne 0 \rightarrow\tilde{e}=0$

$$
f^2 = Y =W =0, \quad \alpha_1 = \alpha_2 \left(\frac{1-q}{q}\right).
$$
The solutions of the equations \eqref{n55}, \eqref{n56} have the form:

\begin{enumerate}

\item
$\alpha_1 = \alpha_2 =0.$ \quad All functions \quad $\eta^{\alpha\beta}\quad $ and functions \quad $\alpha_3 $ \quad are independent field functions.

\item
$\alpha_2 = a = const \ne 0,$
$$
\alpha_1 = a\left(\frac{1-q}{q}\right), \quad \eta^{pq}= \left(\frac{q-1}{aq}\right)\dot{\alpha}_3n^{33} n^{pq}, \quad $$
$$ n^{22}=\frac{(n^{23})^2 +1}{n^{33}},\quad  \eta^{12} = \eta^{13} \frac{n^{23}}{n^{33}}.
$$
The functions \quad $\alpha_3,\quad\eta^{11}, \quad\eta^{13}, \quad n^{23}, \quad n^{33}$ \quad are independent field functions.

\item
$ (q-1) \dot{\alpha}_2 \ne0, \quad \alpha_1 = \alpha_2 \left(\frac{1-q}{q}\right)$
$$
\eta^{pq} = \dot{\alpha}_2 \left(\frac{1-q}{Aq}\right)n^{pq}, \quad n^{23} = -\frac{1}{A\dot{\alpha}_2} \left(A{\dot{A}} + \alpha_2 \dot{\alpha}_2 +\dot{\alpha}_3 A n^{33} \right), \quad $$
$$
n^{22}=\frac{(n^{23})^2 +1}{n^{33}}, \quad
\eta^{12} = \frac{1}{\alpha_2 n^{33}} \left({\dot{A}} + (\alpha_2 n^{23} -A)\eta^{13} \right),.
$$
The functions $A,\quad \alpha_2,\quad \alpha_3,\quad\eta^{11},\quad\eta^{13},\quad n^{33}$ are independent field functions.

\end{enumerate}

\quad

\item \quad $q=-1.$ \quad

\quad

 In this case $\tilde{e}=\frac{1}{e}$ and the solution of equations \eqref{n52}, \eqref{n53} can be represented as:
$$
\eta^{11} = e(\dot{A} + (\alpha_2 n^{23}-A)\eta^{13} -\alpha_2 n^{33}\eta^{12}), \quad n^{22}=\frac{(n^{23})^2 +1}{n^{33}},\quad \eta^{pq}=n^{pq}\exp{\sigma_0},
$$
where
$$
\eta^{12} = \left(\frac{n^{23}}{n^{33}}\right)_{,0} +2e(\dot{A} + \dot{\alpha}_pn^{p3}) - e\alpha_2\exp{\sigma_0} + \eta^{13}\left(\frac{n^{23}}{n^{33}}\right),\quad \eta^{13} =e n^{33}(\dot{\alpha}_1- A\exp{\sigma_0}).
$$
The functions \quad $A,\quad \alpha_\beta,\quad \sigma_0,\quad n^{23},\quad n^{33}$ \quad are independent field functions.
The functions $\quad \eta^{pq}, \quad n^{22}
$ \quad are dependent field functions.
\end{enumerate}

\subsection{\textbf{Spaces}\quad $V^*_4(VII_{n})$}

There are only two types of nonequivalent null homogeneous Petrov spaces of type $V_3(VII)$ (see \cite{3n}-\cite{5n}).

\subsubsection{\textbf{Spaces}\quad $V^*_4(VII_{1})$}

The first type is defined by group operators of the form:
$$
{\rm X}_{a} =\delta_a^1 p_{2}+
\delta_a^2p_{3}+\delta_a^3(p_{1} +u^{3} (2\cos{a} p_{3} -p_{2})+u^{2} p_{3}).
$$
$$
Y_{a} =\delta_a^1 p_{1}+ \delta_a^2(p_{2} \sin p-p_{3} \sin(p+a))\exp q + \delta_a^3(p_{2} \cos{p}-p_{3} \cos(p+a))\exp q,
$$
where:
$p=u^{3} \sin a ,\quad q=u^{3} \cos a.$

\noindent
Nonzero structure constants are:
\begin{equation}\label{n57}
C^2_{12} = C^3_{13} = \cos{a}, \quad C^3_{12} = - C^2_{13} = \sin{a} \Rightarrow
C^{23} = -C^{32} = \cos{a},
\end{equation}
$$
C^{33} = C^{22} = \sin{a},\quad C_a =-2\delta_a^1 \cos{a},
$$
$$
f^1 =0, \quad f^2 =\alpha_2\sin{a} - \alpha_3\cos{a}, \quad f^3 =\alpha_2\cos{a} - \alpha_3\sin{a}.
$$
Obviously, ${f^{2}}^2 + {f^{3}}^2 \ne 0$ (otherwise the electromagnetic field disappears). Without loss of generality, we assume $f^2 \ne 0$. Using. Then Maxwell equations take the form:
\begin{equation}\label{n58}
A=(\dot{\alpha}_1 - y_3f^2 +y_2f^3)\exp{\sigma_0} = 0 ,
\end{equation}
$$
y_1W + \dot{\alpha}_p n^{p2} =\dot{W}_1\cos{a}-\dot{Y}_1\sin{a}, \quad
y_1Y + \dot{\alpha}_p n^{p3} =\dot{W}_1\sin{a}+\dot{Y}_1\cos{a}.
$$
\begin{equation}\label{n59}
y_1 = \frac{1}{W}(\dot{\alpha_2}n^{22} + \dot{\alpha_3}n^{23} + \dot{W}\cos a - \dot{Y}\sin a), \quad y_3 =\frac{1}{f^2}(\dot{\alpha}_1+ y_2f^3).
\end{equation}
After elimination the function $y_1$ from the system of equations \eqref{n58} one has the equation:
\begin{equation}\label{n60}
(\alpha_2\dot{\alpha}_3 - \alpha_3\dot{\alpha}_2 ) +(\alpha_2\dot{\alpha}_2 + \alpha_3\dot{\alpha}_3 )\tan{a} = (Y\dot{W} - W\dot{Y} ) -(W\dot{W} + Y\dot{Y})\tan{a}.
\end{equation}
The functions $\eta^{pq}$ can be expressed through the functions $W, Y$ from the relations \eqref{n27}:
\begin{equation}\label{n61}
\eta^{22}=\frac{{f^2}^2 + W^2}{f^2Y-f^3W}\exp{\sigma_0}, \quad \eta^{23}=\frac{f^2f^3 + YW}{f^2Y-f^3W}\exp{\sigma_0}, \quad \eta^{33}=\frac{{f^3}^2 + Y^2}{f^2Y-f^3W}\exp{\sigma_0},
\end{equation}
Functions $f^p, Y, W$ must be found from equation \eqref{n60}, function $\sigma_0$ is an arbitrary function of $u^0$.
Let us denote:
$$
Y= \exp{\beta}\cos(\varphi+\beta\tan{a}), \quad W = \exp{\beta}\sin(\varphi +\beta\tan{a}),
$$
$$
\alpha_2 = \exp{\gamma}\cos(\omega -\gamma\tan{a}), \quad \alpha_3 = \exp{\gamma}\sin(\omega -\gamma\tan{a}),
$$
where $ \beta, \gamma, \varphi, \omega$ are functions of the variable $u^0$. Then equation \eqref{n60} takes the form:
$$
\dot{\omega}\exp{2\gamma}= (Y\dot{W} - W\dot{Y} ) -(W\dot{W} + Y\dot{Y})\tan{a}.
$$
Let's find all solutions of this equation.

\quad

\noindent
1) $\dot{\omega}=0 \Rightarrow $
\begin{equation}\label{n62}\left\{\begin{array}{ccc} {W= \exp{\beta}\sin(\beta\tan{a}), \quad Y= \exp{\beta}\cos(\beta\tan{a}),}
\\\\{\alpha_2 = \exp{\varphi}\sin(\varphi\tan{a}), \quad \alpha_3 = \exp{\varphi}\cos(\varphi\tan{a})\Rightarrow }\\\\{f^2=-\exp{\varphi}\cos(\varphi\tan{a}+a), \quad f^3=\exp{\varphi}\sin(\varphi\tan{a}+a).} \end{array}\right.
\end{equation}

\quad

\noindent
2) $\dot{\omega}\ne 0 \Rightarrow $

\begin{equation}\label{n63}
 \left\{\begin{array}{ccc} {W= \exp{\beta}\sin(\varphi\tan{a}),\quad Y= \exp{\beta}\cos(\varphi\tan{a}),}
 \\\\
 { \alpha_2 = \exp{\varphi}\left(\frac{\dot{\beta}-\dot{\varphi}\tan{a}}{\dot{\omega}}\right)^{\frac{1}{2}}\cos\Omega,
 \quad\alpha_3 = \exp{\varphi}\left(\frac{\dot{\beta}-\dot{\varphi}\tan{a}}{\dot{\omega}}\right)^{\frac{1}{2}}\sin\Omega,}
  \\\\{f^2=\exp{\varphi}\left(\frac{\dot{\beta}-\dot{\varphi}\tan{a}}{\dot{\omega}}\right)^{\frac{1}{2}}
\cos{(\Omega-a)},\quad f^3=\exp{\varphi}\left(\frac{\dot{\beta}-\dot{\varphi}\tan{a}}{\dot{\omega}}\right)^{\frac{1}{2}}
\sin{(\Omega-a)}}, \end{array}\right.
\end{equation}
where
$$
\quad \Omega = \omega-\left(\varphi +\frac{1}{2}\ln\frac{\dot{\beta}-\dot{\varphi}\tan{a}}{\dot{\omega}}\right)\tan{a}
$$
For the both cases, the functions $\eta^{11}, \eta^{13}$ have the form:
$$
\eta^{11} = \frac{1}{W}(\dot{\alpha_2}n^{22} + \dot{\alpha_3}n^{23} + \dot{W}\cos a - \dot{Y}\sin a), \quad \eta^{13} =\frac{1}{f^2}(\dot{\alpha}_1+ \eta^{12}f^3).
$$
Functions\quad $\alpha_1, \quad\eta^{12},\quad \beta, \quad \varphi, \quad \omega, \quad \sigma_0.$ are the independent field functions.

\subsubsection{\textbf{Spaces}\quad $V^*_4(VII_{2})$}

The  group operators have the form:
$$
{\rm X} _{a} =\delta_a^1{\rm p}_{1} + \delta_a^2{\rm p}_{1}\frac{\exp - q}{\cos p} +{\rm p}_{2} \frac{\cos \left(p-\alpha \right)}{\cos p}+\delta_a^3 ({\rm p}_{3} +u^{2} {\rm X}_{2}) .
$$
$$
Y_{1} =p_{1} ,\quad Y_{2} =p_{2}\sin a\frac{\exp q}{\cos p}-p_{1}\frac{\sin p}{\cos p} , \quad Y_{3} =p_{3} +u^{1} (p_{1}\cos a -Y_{2}\sin a),
$$
where denoted:
$p=u^{3} \sin a ,\quad q=u^{3} \cos a \quad (a$ is a parameter =const)

\noindent
Nonzero structure constants:
$$
 C^1_{13} = C^2_{23} = \cos{a}, \quad C^1_{23} = - C^2_{13} = \sin{a} \Rightarrow c_a =2\delta_a^3 \cos{a},
$$
 $$
 C^{21} = -C^{12} = \cos{a}, \quad C^{11} = C^{22} = \sin{a},
 $$
 $$
 f^1 =\alpha_2\cos{a}+ \alpha_1\sin{a},\quad f^2 =\alpha_2\sin{a} - \alpha_1\cos{a}, \quad f^3 =0.
 $$
Let us consider two options separately: $\dot{\alpha_2}=0,$ and $\dot{\alpha_2}\ne 0.$

\begin{enumerate}

\item $\dot{\alpha}_2=0 $

\quad

Obviously in this case \quad $\alpha_2 = e =const \ne 0 \Rightarrow \quad \alpha_1 = e \tan{a}, \quad \alpha_3 =e\beta(u^0). $

The Maxwell equations take the form:
\begin{equation}\label{n64}\left\{\begin{array}{cc} {A=0, \quad(y_3n^{23} -y_2n^{33})\tan{a} + \dot{\beta}n^{33}\sin{a} =\exp\sigma_0,}
\\\\
{(y_3n^{23} -y_2n^{33}) + \tan{a}\exp\sigma_0}=0.
\end{array}\right.
\end{equation}
The solution can be represented as:
$$
\eta^{pq}=\exp{\sigma_0}n^{pq}= (\dot{\beta}n^{33}\sin{a} \cos^2{a})n^{pq}, \quad \eta^{12}=\frac{n^{23}\eta^{13} + \dot{\beta}n^{33}\sin^2{a} \cos{a}}{n^{33}}, \quad
$$

$$\alpha_2 = e =const \quad \alpha_1 = e \tan{a}, \quad \alpha_3 =e\beta.
$$

$$
\exp\sigma_0= \dot{\beta}n^{33}\sin{a} \cos^2{a},  \quad n^{22}=\frac{(n^{23})^2 +1}{n^{33}}
$$
The functions \quad $\beta, \quad\eta^{11},\quad\eta^{13} \quad n^{23}, \quad n^{33}.$ \quad are independent field functions.

\quad

\item $\dot{\alpha}_2 \ne 0.$

\quad

Let us introduce the following functions.
$$P=y_2 \dot{\alpha}_2 + y_3\dot{\alpha}_3, \quad P_1 = \frac{\alpha\dot{\alpha} + \alpha_2\dot{\alpha_2}\sin^{2}{a} + \alpha^2 y_3 \cos{a}}{\alpha_2 \alpha \cos{a}}, \quad P_2 = \frac{\alpha\dot{\alpha} + \alpha_2\dot{\alpha_2}}{ \alpha \cos^{2}{a}}.
$$
Maxwell equations can be represented in the following form:
\begin{equation}\label{n65}\left\{\begin{array}{cccc} {y_2n^{33} -y_3 n^{23}=\frac{1}{\alpha_2}\left(\dot{A}+ Ay_3\cos a + \frac{\alpha_2\sin a}{\cos a}\exp\sigma_0\right),}
\\\\{\dot{\alpha}_2 n^{23} + \dot{\alpha}_3 n^{33} =\frac{1}{\cos a}\left(\dot{A} + \frac{\alpha_2}{\sin a \cos a}\exp\sigma_0\right),}\\\\{Y = 0,  \quad A \exp\sigma_0 =\dot{\alpha_1} \Rightarrow}\\\\{f^2 = f^3 =W =0 \quad \Rightarrow \alpha_1 = \alpha_2 \tan a, \quad  f^1 =\frac{\alpha_2}{\cos a}.} \end{array}\right.
\end{equation}
The solutions of the system of equations \eqref{n65} have the form:

\quad

\begin{enumerate}

\item  $ P \ne 0,\quad \alpha_1=\alpha_2 \tan{a}, $

$$
n^{23}= \frac{(P_1 \dot{\alpha_2} + P_2y_3)}{P}, \quad n^{33}= \frac{(P_2y_2 - P_1 \dot{\alpha_3})}{P},\quad \exp\sigma_0 =\frac{\dot{\alpha_2}\sin{a}}{\alpha} \Rightarrow
$$
$$
\eta^{23}= \frac{\dot{\alpha_2}(P_1 \dot{\alpha_2} + P_2\eta^{13})\sin{a}}{\alpha P}, \quad
\eta^{33}= \frac{\dot{\alpha_2}(P_2\eta^{12} - P_1 \dot{\alpha_3})\sin{a}}{\alpha P}
$$
where it is denoted \quad $\alpha = A\cos{a}$.\quad

Functions\quad $\alpha, \quad\alpha_2, \quad \alpha_3, \quad\eta^{1\alpha}.$ \quad are the independent field functions.

\quad

\item $ P = 0 \Rightarrow y_2 =-y_3\frac{\dot{\alpha_3}}{\dot{\alpha_2}}.$

\quad

From the equations \eqref{n65} it follows:
$$
\dot{\alpha}_2 P_1 + y_3 P_2 = 0, \quad \dot{\alpha_3}n^{33}+\dot{\alpha_2 n^{23}}=P_2.
$$
The solution has the form:
$$
n^{23} = \frac{\dot{\alpha}\alpha + \dot{\alpha}_2\alpha_2}{\alpha \cos^2{a}} -n^{33}\frac{\dot{\alpha}_3}{\dot{\alpha}_2}, \quad \exp{\sigma_0}= \frac{\dot{\alpha}_2}{\alpha}\sin{a}\Rightarrow
$$
$$ \eta^{23} = \frac{\dot{\alpha}_2(\dot{\alpha}\alpha + \dot{\alpha}_2\alpha_2)\sin{a}}{(\alpha \cos{a})^2} -\eta^{33}\frac{\dot{\alpha}_3}{\dot{\alpha}_2}, \quad \eta^{22}=\frac{(\eta^{23})^2 +(\frac{\dot{\alpha}_2}{\alpha}\sin{a})^2 }{\eta^{33}},
$$
$$
\eta^{12}=y_2 = \frac{\dot{\alpha}_3(\alpha\dot{\alpha}+\alpha_2\dot{\alpha}_2\sin^2{a})}
{\alpha(\dot{\alpha}\alpha_2+\dot{\alpha}_2\alpha\cos^2{a})},\quad
\eta^{13}=y_3 =- \frac{\dot{\alpha}_2(\alpha\dot{\alpha}+\alpha_2\dot{\alpha}_2\sin^2{a})}
{\alpha(\dot{\alpha}\alpha_2+\dot{\alpha}_2\alpha\cos^2{a})}.
$$

The functions \quad $\alpha, \quad\alpha_2, \quad \alpha_3, \quad\eta^{11}$ \quad are independent field functions.

\end{enumerate}

\end{enumerate}

\section{Conclusion}

Thus, the problem of completely classifying exact solutions of Maxwell equations in homogeneous Petrov spaces has been solved. All nonequivalent solutions for null homogeneous Petrov spaces with non-solvable groups of motions $G_3(VIII), G_3(IX)$ are found in \cite{60n}. For non-null homogeneous Petrov spaces, the problem has been solved in \cite{56n} - \cite{58n}.
Consequently, when integrating the Einstein-Maxwell equations, only the Einstein equations remain to be considered. Note that all the obtained exact solutions of Maxwell vacuum equations in homogeneous Petrov spaces are explicitly expressed in terms of arbitrary independent functions.
Therefore, in all homogeneous Petrov spaces, the Einstein-Maxwell equations reduce to systems of ordinary differential equations involving only these independent functions.

\end{document}